\documentclass[12pt]{article}
\usepackage{epsfig}
\usepackage{amssymb}
\usepackage{pifont}
\usepackage{cite}
\usepackage{here}

\begin{document}

\newcommand{\rhobar}{\bar {\rho}}
\newcommand{\etabar}{\bar{\eta}}
\newcommand{\epsilonk}{\left|\varepsilon_K \right|}
\newcommand{\vubovcb}{\left | \frac{V_{ub}}{V_{cb}} \right |}
\newcommand{\vubsvcb}{\left | V_{ub}/V_{cb}  \right |}
\newcommand{\vtdovts}{\left | \frac{V_{td}}{V_{ts}} \right |}
\newcommand{\epsp}{\frac{\varepsilon^{'}}{\varepsilon}}
\newcommand{\dmd}{\Delta m_d}
\newcommand{\dms}{\Delta m_s}

%%%%%%%%\psdraft %non mette le figure
%%%%%%%%%%%%%%%%%%%%%%%%%%%%%%% titlepage %%%%%%%%%%%%%%%%%%%%%%%%%%%%%%%%%%%%
%  \docnum{CERN--EP/YY--NNN}
%  \date{someday}

  \title{Inferring $\rhobar$ and $\etabar$ of the CKM matrix \\
   -- A simplified, intuitive approach --}

\author{G. D'Agostini\\
 \mbox{\it Universit\`a di Roma ``La Sapienza''} \\
 \mbox{\it and Sezione INFN di Roma1, Roma, Italy.}
}

%\Instfoot{xx}{
%        Part of lectures given a the VI LNF Spring School 
%        ``Bruno Touscheck'', May 2001. \\  
%        {\rm Email}: {\tt dagostini@roma1.infn.it}\\
%        {\rm URL}: {\tt http://www-zeus.roma1.infn.it/$^\sim$agostini}}         
\date{}

\maketitle

\begin{abstract} 
This analysis is based on the same ideas 
and numerical inputs of the recent paper by Ciuchini et al. on the
subject.  Some approximations are applied,
which make analytical calculations applicable in most of the work, 
thus avoiding Monte Carlo integration.
The final result is practically identical to the one 
obtained by the more detailed numerical analysis.
\footnote{Based on the series of lectures ``Statistical methods for
frontier physics'' given at the VI LNF Spring School 
        ``Bruno Touscheck'', Frascati, Italy, May 2001. A version of the
     paper with original {\it Mathematica} eps figures (higher quality,
     but too large for the arXive) is available at the author's URL.  

     {\rm Email}: {\tt dagostini@roma1.infn.it}.
     {\rm URL}: {\tt http://www-zeus.roma1.infn.it/\,$\tilde{ }$\,agostini.}
} 
\end{abstract}

\section{Introduction}
The ``2000 CKM-triangle analysis'' by Ciuchini et al.~\cite{GOR}
is in my opinion the most accurate way 
to use consistently all pieces of experimental and theoretical
information relevant to infer, within the
framework of the Standard Model (SM), the parameters of the 
unitarity triangle. The basic idea is that 
beliefs (i.e. probabilities) 
on the value of each {\it input} quantity are propagated into beliefs 
about the {\it output} quantities, namely   $\rhobar$ and $\etabar$, 
and SM related quantities. The propagation is performed using the rules 
of logic applied to uncertain events, i.e. probability theory, 
without arbitrariness or ``prescriptions''. All hypotheses are 
clearly stated and the influence on the results of 
\underline{reasonable} variations of models and model parameters 
have been checked and found irrelevant (see Ref.~\cite{GOR}
for details). Such a propagation of uncertainty, done by 
responsible people on physics quantities they feel
knowledgeable about should not be confused with the kind of 
mathematical games described in Appendix A of Ref. \cite{Orsay}. 
In  other words, when we say that the value of a quantity
has 50\% probability to lie in a certain region, \underline{given}
well defined hypotheses and with reasonably large flexibility of them,
 means that we are equally confident that it could be there or elsewhere.
This is in contrast to 
``95\% C.L. results'' which should not be interpreted 
as 95\% confidence on a certain 
statement (see e.g. Section 5.2 of Ref. \cite{Maxent98}, and 
the Conclusion of Ref. ~\cite{Read} for an independent 
account). 

The motivation of this note is to illustrate a simplified  
way to perform the analysis of Ref.~\cite{GOR}. In fact, 
the Monte Carlo integration used there could seem
a bit mysterious. 
As well known, approximations usually loose accuracy,
 but often help in gaining awareness. And this is exactly the spirit
of this note. Therefore, I have  little to add here to 
phenomenology, experimental data and inferential framework 
of the analysis, and I point
to Ref. ~\cite{GOR} and references therein.  

The note is organized in the following way. 
In Section \ref{sec:constraints} the relations between 
CKM parameters are schematically recalled. 
In the following sections 
the reweighting effect on the points of the 
$\rhobar$--$\etabar$ plane due to
all pieces of information, is shown in detail.
The final result is then compared to that of  Ref.~\cite{GOR}.

\section{Uncertain constraints}\label{sec:constraints}
The relations (1)--(4) of  Ref.~\cite{GOR} can be written 
in the following way 
\begin{eqnarray*}
 \vubovcb\hspace{0.92cm}\Rightarrow\hspace{3.0cm}
\rhobar^2 + \etabar^2 & = & a \hspace{3.7cm} (C_1) \\
\dmd \hspace{0.92cm}\Rightarrow\hspace{2.0cm} 
(1-\rhobar^2)+\etabar^2 & = & b \hspace{3.7cm}(C_2)  \\
\epsilonk \hspace{1.1cm}\Rightarrow\hspace{1.6cm} 
\etabar\left[1+c\,(1-\rhobar)\right] &=& d  \hspace{3.7cm} (C_3) \\
 (\mbox{mainly}) \  \dms \hspace{1.0cm}\Rightarrow\hspace{2.0cm} 
(1-\rhobar^2)+\etabar^2 & = & \frac{e}{\dms}  
\hspace{3.0cm}  (C_4)
\end{eqnarray*}
where $a$,$b$, \dots $e$ are the constraint parameters, functions 
of many quantities; the most crucial of those determined experimentally
 are indicated on the left hand side of the equations. 
Note that the order of $C_3$ and $C_4$ is exchanged with respect to 
Eqs. (3)--(4) of  Ref.~\cite{GOR}. This is because,
since the present information concerning $\dms$ 
is of different quality with respect to the other quantities, 
the constraint $C_4$ needs a more careful treatment than the others,
and it will be introduced after $C_1$--$C_3$.  
This is also the reason why $\dms$ appears explicitly in 
the constraint $C_4$.

In the ideal case all parameters are perfectly known, 
and the constraints would give curves in the $\rhobar$--$\etabar$ plane. 
For example, $C_1$ would give a circle of radius $\sqrt{a}$. 
In other terms, due to this constraint all 
points of the circumference would be appear to us likely, 
\underline{unless}
there is any other experimental piece of information
(or theoretical prejudice) to assign a different weight to 
different points. In the ideal case 
 the p.d.f. describing our beliefs 
in the $\rhobar$ and $\etabar$ values would be 
\begin{equation}
f (\rhobar, \etabar\,|\,a) = \delta(\rhobar^2 + \etabar^2 - a)\,,
\end{equation}
with the Dirac delta meaning the limit 
of a very narrow p.d.f. concentrated on the circumference. 
Similar arguments hold for the parameters of the
other constraints. Let us call them generically $p$ 
(the fact that $C_3$ depends on two parameters is conceptually
irrelevant). 

In the real case $p$ itself is not perfectly known. 
There are values which are more likely, and values 
which are less likely, classified
by the p.d.f. $f(p)$. This means that, e.g. for the  parameter $a$,
 we deal with an infinite number 
of circles, each having its weight $f(a)$. It follows that the points
of the ($\rhobar$, $\etabar$) plane get different weights. 

The probability theory teaches us how to evaluate $f(\rhobar, \etabar)$
taking into account all possible values of $p$:
\begin{equation}
f(\rhobar, \etabar) = \int f(\rhobar, \etabar\,|\,p)\,f(p)\,\mbox{d}p\,. 
\end{equation}
Now the problem becomes how to evaluate $f(p)$, knowing that 
each $p$ depends on many input quantities $x_i$, 
denoted all together with 
$\mathbf{x}$ and described, in the most general case, 
 by a joint p.d.f. $f(\mathbf{x})$.
In most 
 cases of practical interest, including the one we are discussing, 
each $x_i$ can be considered independent from the others, 
and the joint distribution simplifies to  
 $f(\mathbf{x})\approx \prod f(x_i)$. Calling $p(\mathbf{x})$ the function  
which relates $\mathbf{x}$ to $p$, $f(p)$ can be generally
 obtained as 
\begin{equation}
f(p) = \int f(\mathbf{x})\,\delta(p-p(\mathbf{x}))\,\mbox{d}\mathbf{x}\,.
\end{equation}
This integral, as well as the previous integrals, is usually performed
by Monte Carlo techniques. 
The calculations can be simplified in the following way. 
\begin{itemize}
\item
First, we rely on the central limit theorem, assuming $f(p)$ 
to be Gaussian
for all parameters (but with no constraint on the shape of 
$f(\rhobar, \etabar)$\,!). 
\item
Expected value, standard (deviation) uncertainty and correlation
are basically obtained by the usual propagation (see Ref.~\cite{conPeppe}
for a similar application). 
\item
Non linear effects in the propagation have been taken into
account up to second order, using formulae of Ref.~\cite{conMirko}
(see this paper for the practical modeling of uncertainties and 
treatment  of asymmetric cases).
\item
The correlation between $c$ and $d$ has been taken into account 
building a bi-variate $f(c,b)$. 
\end{itemize}

As input quantities, Table 1 of Ref.~\cite{GOR} is used, combining 
properly (i.e. quadratically) the standard deviations. 
For example, for $\sigma\left(f_{B_d} \sqrt{\hat B_{B_d}}\right)$ 
expressed  in MeV we obtain 
$\sqrt{25^2+(20/\sqrt{3})^2} = 27.5 $, where 
$20/\sqrt{3}$ stands for the standard deviation 
of a uniform distribution of half width 20 MeV.   
Note that, contrary to what some authors critical 
about Ref.~\cite{GOR} (and the related papers and presentations  
to conferences) think, I have the impression that my colleagues 
tend to make slightly conservative assessments
of uncertainties. This feeling that I had {\it a priori}
discussing with them is somewhat confirmed
{\it a posteriori} by the excellent overlap of the partial
inference by each constraint (as it will be shown in 
Fig.~\ref{fig:contours}) and self-consistency 
between input parameters and values coming out of
the inference obtained without the their contribution
(see  Ref.~\cite{GOR}).  

The following results are obtained in terms of expected values and
standard deviations (``$p=\mbox{E}(p)\pm\sigma(p)$''): 
\begin{eqnarray}
a &=&  0.145 \pm 0.031 \\
b &=&  0.73 \pm  0.19  \\
c &=&  3.4  \pm  0.9   \\
d &=&  1.23 \pm  0.33  \hspace{1.0cm} [\,\rho(c,d) = 0.76\,] \\
e &=&  12.7 \pm  1.2 \,\mbox{ps}^{-1} 
\end{eqnarray}
Only the correlation between $c$ and $b$ is relevant, and all
others can be neglected, being below the 10\% level 
(even that between $b$ and $e$, related by $\dmd$, is negligible, 
being only +7\%). The radii of the circles given by $C_1$ 
and $C_2$ are $0.38\pm 0.04$ and $0.85\pm0.11$, respectively. 
These radii have the meaning of the sides of the unitarity triangle opposite
to $\beta$ and $\gamma$, respectively, provided
 by $C_1$ and $C_2$ alone. 

\newpage

\section{Partial results from $C_1$, $C_2$ and $C_3$}

\subsection{ $\vubovcb$}
\begin{figure}[b!]
\centering\epsfig{file=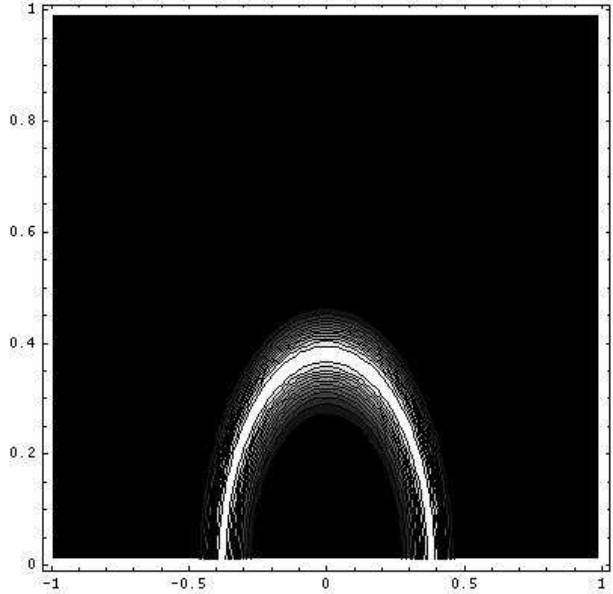,clip=,width=8.0cm}
%\mbox{}\vspace{-13.0cm}ciao\mbox{}\hspace{1.0cm} {\large $\eta$}
%\vspace{+13.0cm}
\caption{\small \sf Contour plot of the p.d.f. of Fig. \ref{fig:fC1full} for 
$\etabar \ge 0$. Note that contours are simply iso-p.d.f. levels obtained 
by 12 contour lines. In order to assign them a probabilistic meaning, 
one needs to evaluate the p.d.f. integrals inside the contours.}
\label{fig:fC1contour}
\end{figure}

\begin{figure}
\begin{center}
\begin{tabular}{|c|}
\hline
\epsfig{file=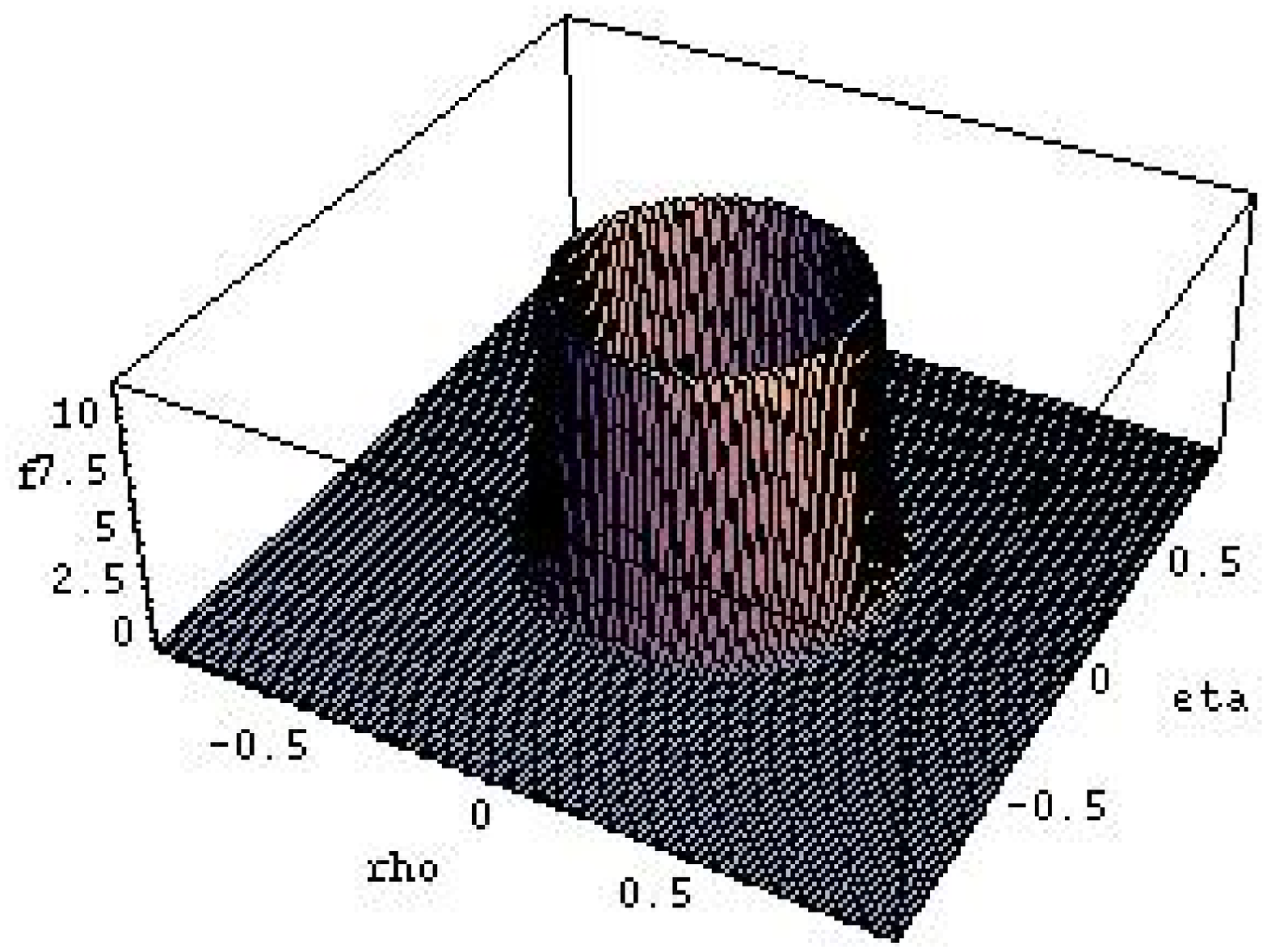,clip=,width=12.0cm} \\
\hline
\epsfig{file=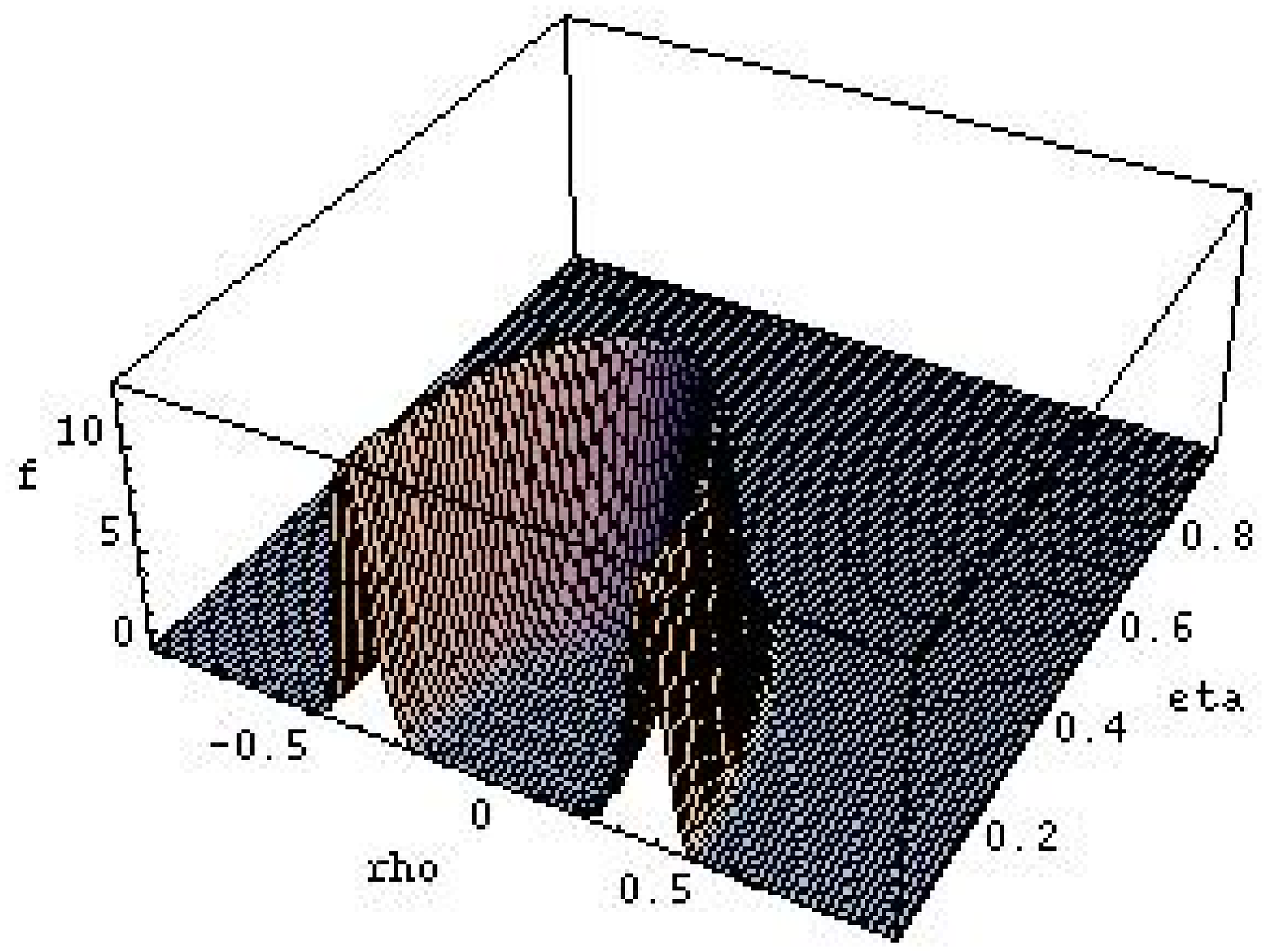,clip=,width=12.0cm} \\
\hline
\end{tabular}
\end{center}
\caption{\small \sl Probability density function of $\rhobar$ and $\etabar$
obtained by the constraint given by  $\vubovcb$.}
\label{fig:fC1full}
\end{figure}

Applying our reasoning to the constraint given by  $\vubovcb$, we obtain
\begin{eqnarray}
f(\rhobar,\etabar\,|\,C_1) &=& \int_{-\infty}^{+\infty}\!
     \delta(\rhobar^2+\etabar^2-a)\,\frac{1}{\sqrt{2\,\pi}\,0.031}\,
     \exp{\left[-\frac{(a-0.145)^2}{2\,(0.031)^2}\right]}\, \mbox{d}a \\
   &=& \frac{1}{\sqrt{2\,\pi}\,0.031}\,
     \exp{\left[-\frac{(\rhobar^2+\etabar^2-0.145)^2}{2\,(0.031)^2}\right]}\,.
\end{eqnarray}
The contour plot shown in Fig. \ref{fig:fC1contour} for $\etabar \ge 0$.
3-D plots are given in Fig. \ref{fig:fC1full} (in all figures 
``rho'' and ``eta'' stand for $\rhobar$ and $\etabar$). 
Note that the plot of Fig. \ref{fig:fC1contour} is obtained slicing 
$f(\rhobar,\etabar\,|\,C_1)$ into 12 iso-p.d.f. contours, and hence
the regions shown there have no straightforward
 probabilistic interpretation,
since one should make the integrals of the p.d.f. inside the region.   
We leave it as an exercise for the interested readers.

\subsection{$\dmd$}
\begin{figure}
\begin{center}
\begin{tabular}{|c|}
\hline
\epsfig{file=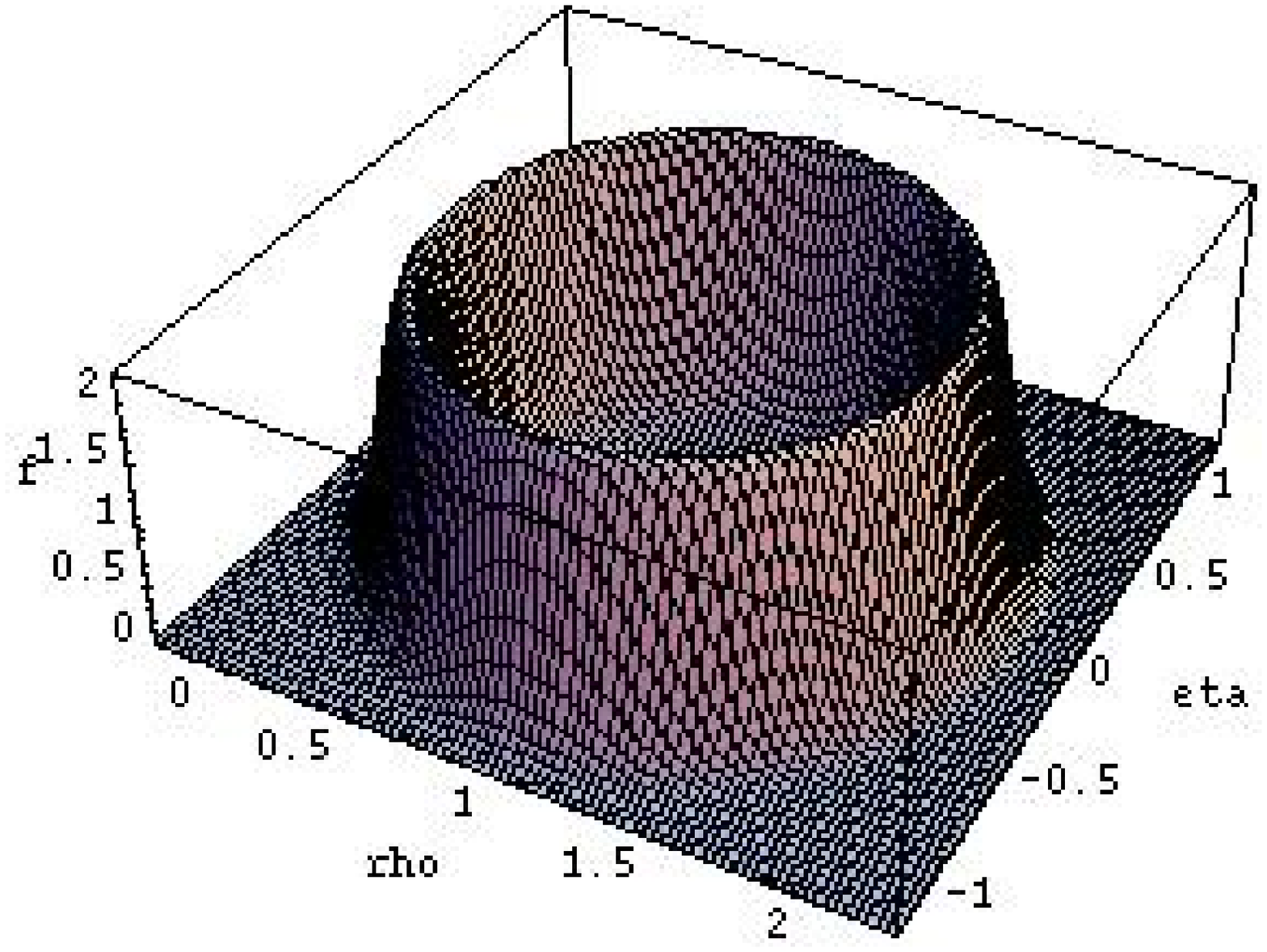,clip=,width=12.0cm} \\
\hline
\epsfig{file=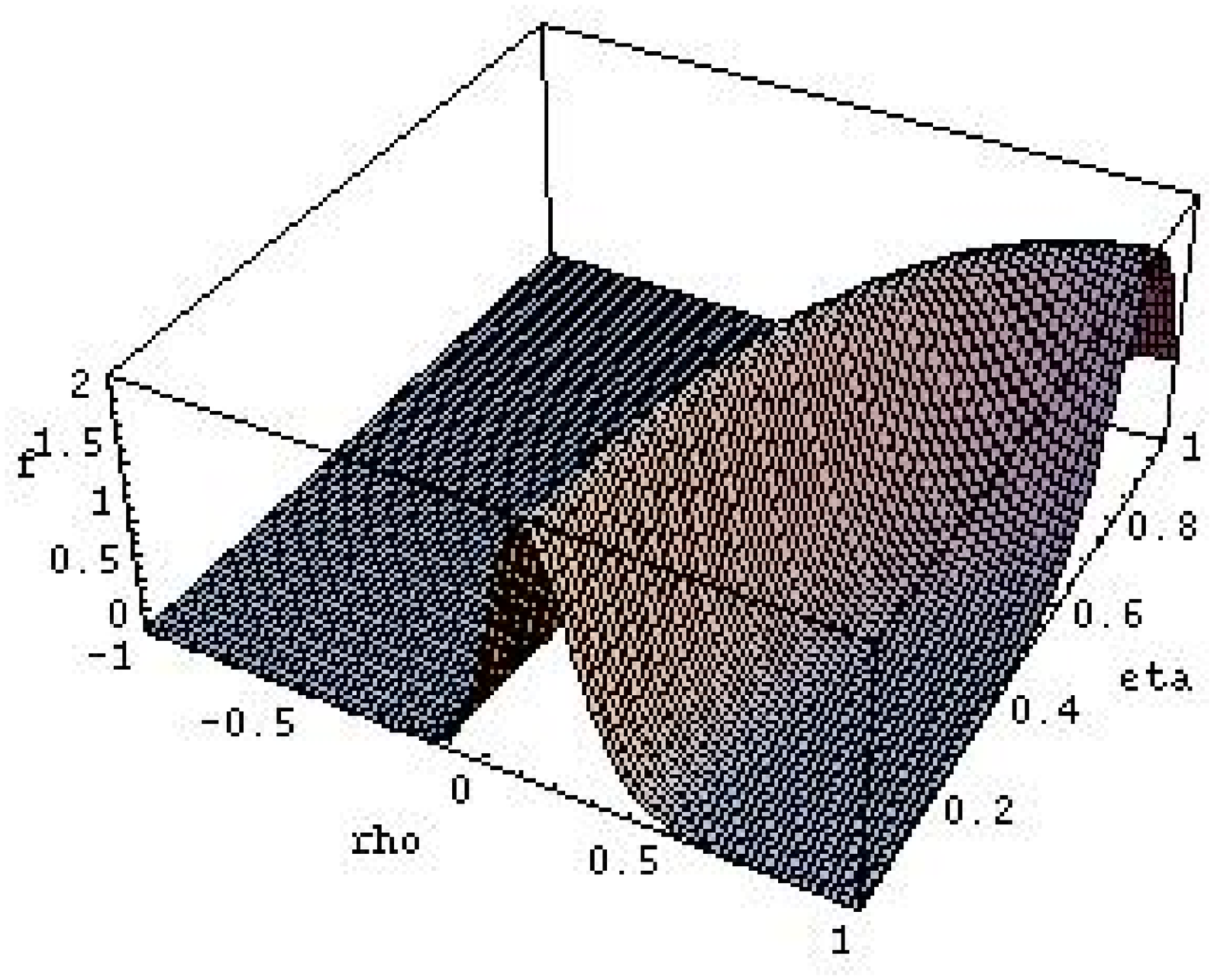,clip=,width=12.0cm} \\
\hline
\end{tabular}
\end{center}
\caption{\small \sl Probability density function of $\rhobar$ and $\etabar$
obtained by the constraint given by $\dmd$ .}
\label{fig:fC2full}
\end{figure}

From the $\dmd$ we get
\begin{eqnarray}
f(\rhobar,\etabar\,|\,C_2) &=& \int_{-\infty}^{+\infty}\!
     \delta((1-\rhobar)^2+\etabar^2-b)\,\frac{1}{\sqrt{2\,\pi}\,0.19}\,
     \exp{\left[-\frac{(b-0.73)^2}{2\,(0.19)^2}\right]}\, \mbox{d}b \\
   &=& \frac{1}{\sqrt{2\,\pi}\,0.19}\,
     \exp{\left[-\frac{((1-\rhobar)^2+\etabar^2-0.73)^2}{2\,(0.19)^2}\right]}\,,
\end{eqnarray}
shown by 3-D plots in Fig. \ref{fig:fC2full}. 

\begin{figure}[b!]
\centering\epsfig{file=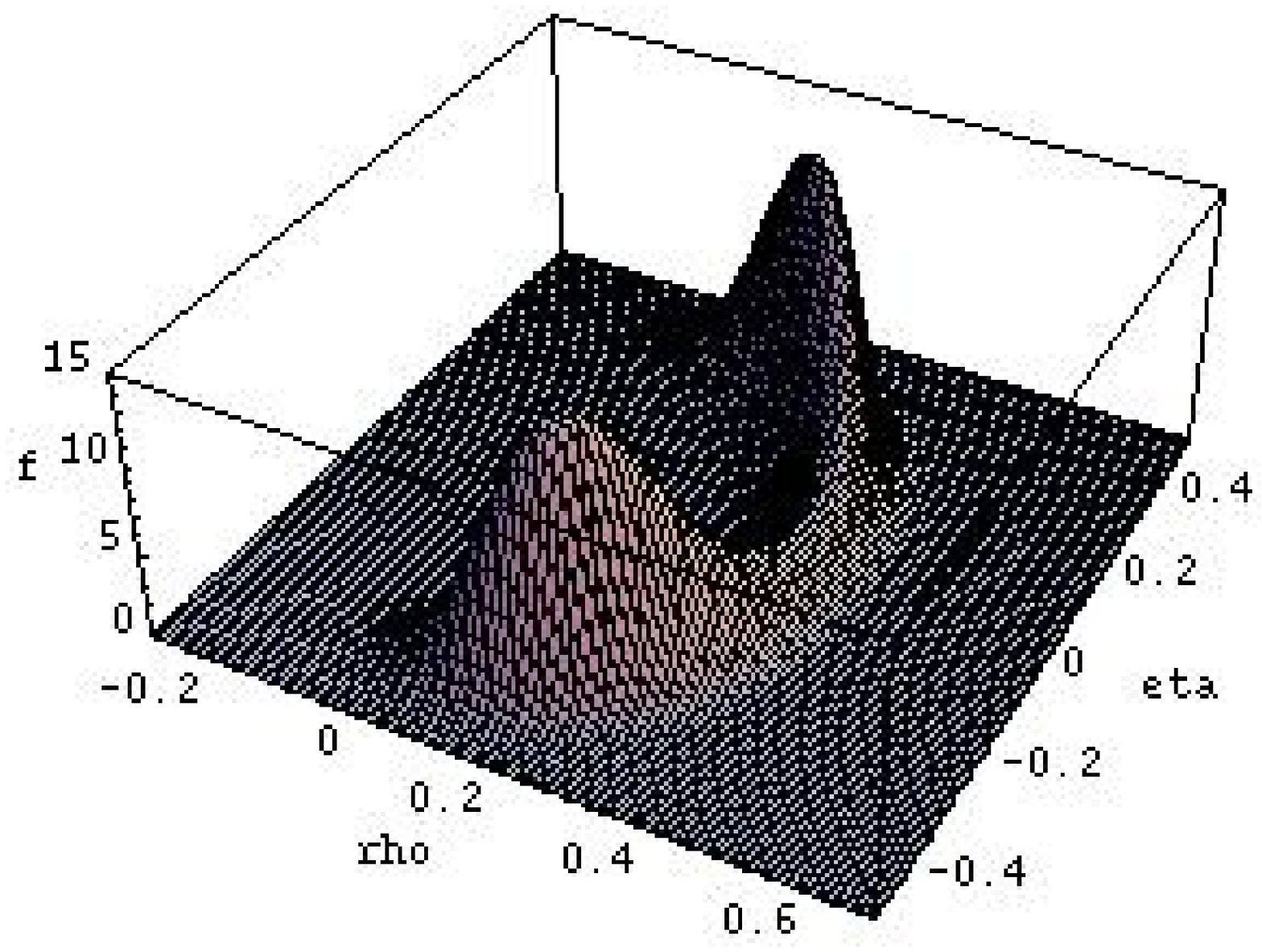,clip=,width=12.0cm}
\caption{\small \sf $f(\rhobar, \etabar)$ from the constraints given
by   $\vubovcb$ and  $\dmd$  .}
\label{fig:fC1C2}
\end{figure}

We can already combine the information from the two constraints, 
just remembering
a condition  stated  in the second paragraph of 
section \ref{sec:constraints}: 
In the ideal case we would consider the points of the circle equally likely 
\underline{if} there were no other experimental or theoretical 
a priori information
which would lead to assign different weights to the 
different points. But this is exactly
what each partial inference does. Therefore, in order to combine the two 
constraints we just need to multiply the 
weights of each point and normalize the distribution: 
\begin{equation}
f(\rhobar, \etabar\,|\,C_1,C_2) \propto 
f(\rhobar, \etabar\,|\,C_1)\, f(\rhobar, \etabar\,|\,C_2). 
\end{equation}
The 3-D plot of the normalized distribution is shown in Fig. \ref{fig:fC1C2}.
We see that with these two constraints only there is still sign ambiguity, 
and, in particular, the distribution is specular with respect to 
the $\etabar=0$ axis.  

\subsection{$\epsilonk$}
\begin{figure}
\begin{center}
\epsfig{file=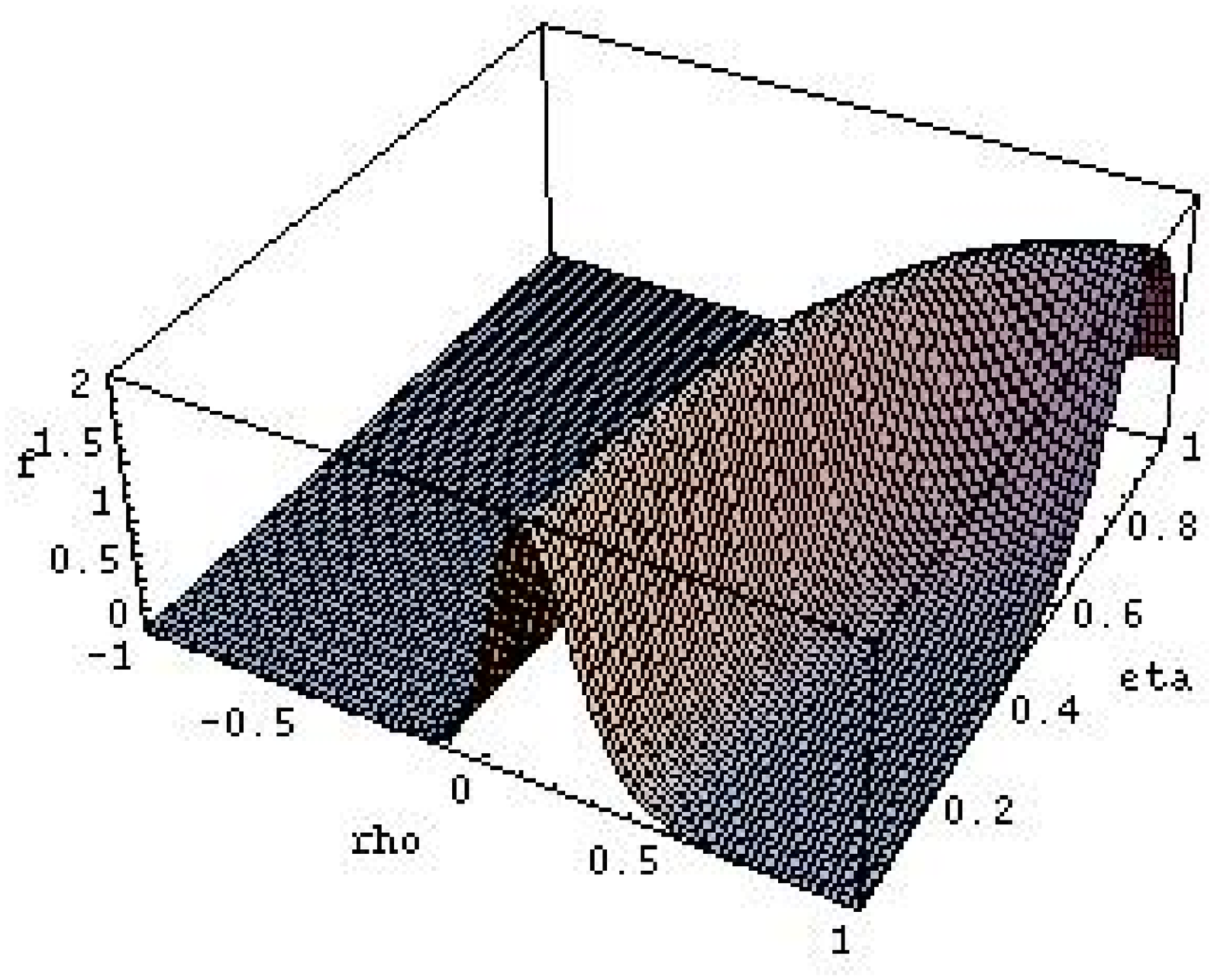,clip=,width=12.0cm} 
\end{center}
\caption{\small \sl  Probability density function of 
$\rhobar$ and $\etabar$
obtained by the constraint given by $\epsilonk$.}
\label{fig:fC3}
\end{figure}

Since the third constraint depends on two strongly correlated
parameters, we need to consider a joint bivariate Gaussian distribution:
\begin{eqnarray}
f(c,d) &=&
\frac{1}{2\,\pi\,\sigma_c\,\sigma_d\,\sqrt{1-\rho_{cd}^2}}\cdot  
  \exp \left\{
                -\frac{1}{2\,(1-\rho_{cd}^2)} 
                 \left[  \frac{(x-\mu_c)^2}{\sigma_c^2} \right.\right.
  \nonumber \\
 && \left.\left.   - 2\,\rho_{cd}\,\frac{(c-\mu_c)(d-\mu_d)}{\sigma_c\,\sigma_d}
        + \frac{(d-\mu_d)^2}{\sigma_d^2}
                 \right]
         \right\} \,,
\label{eq:bivar}
\end{eqnarray}
with $\mu_c=\mbox{E}[c]=3.4$, $\sigma_c=0.9$, 
 $\mu_d=\mbox{E}[d]=1.23$, $\sigma_d=0.33$, and $\rho_{cd}=\rho(c,d)= 0.76$.
The integral 
\begin{equation}
f(\rhobar,\etabar\,|\,C_3) = \int\!\!\!\!\int_{-\infty}^{+\infty}\!
\delta\left(\etabar\,[1+c\,(1-\rhobar)]-d\right) \, 
f(c,d)\,\mbox{d}c\,\mbox{d}c
\end{equation}
can be still evaluated 
analytically but we omit here the final formula, just
giving the 3-D plot of the 
resulting  p.d.f. normalized in the region of interest (Fig. \ref{fig:fC3}).   

\section{Partial combinations of results ($C_1$, $C_1$ and $C_3$)}

\begin{figure}
\begin{center}
\begin{tabular}{|c|}
\hline
\epsfig{file=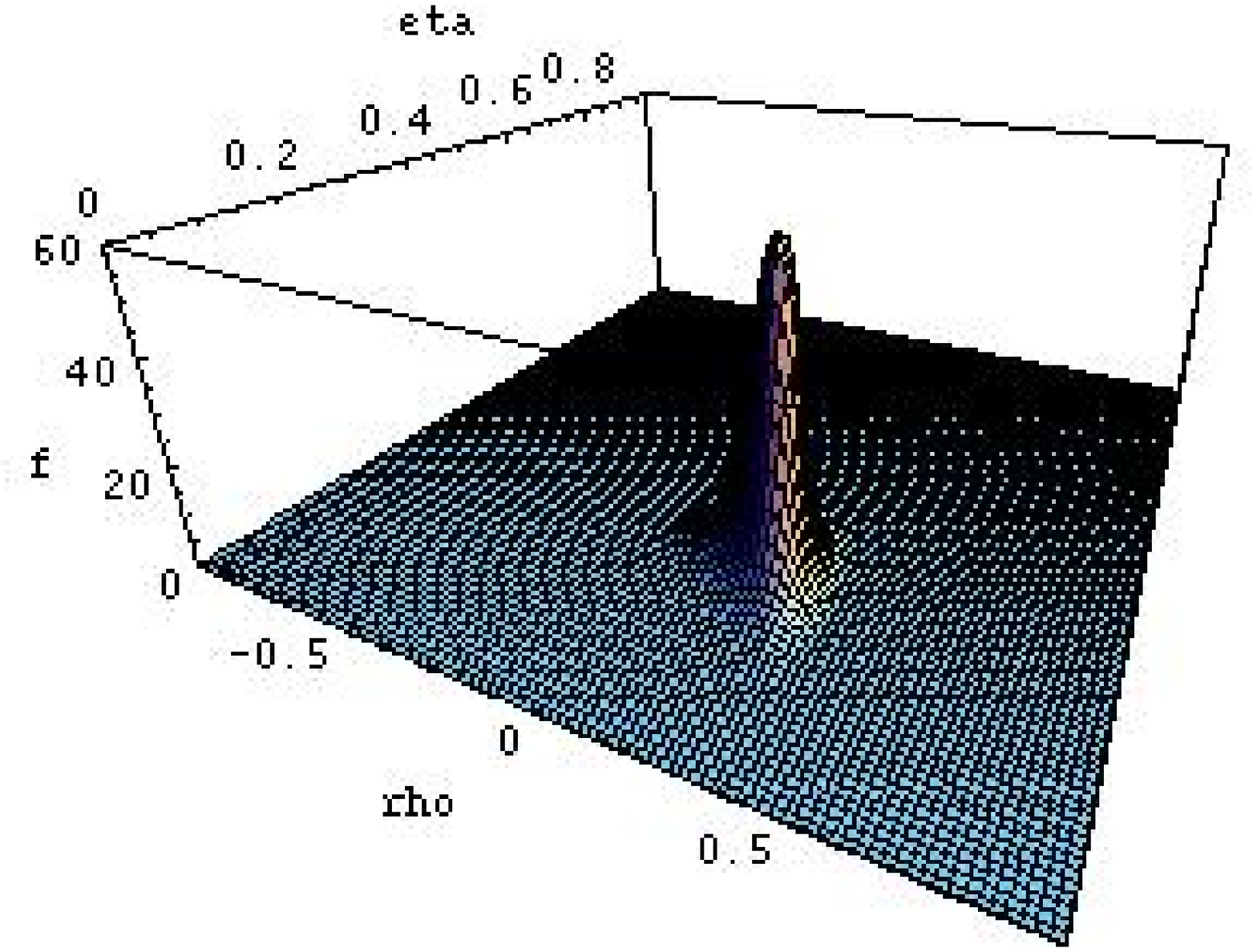,clip=,width=12.0cm} \\
\hline
\epsfig{file=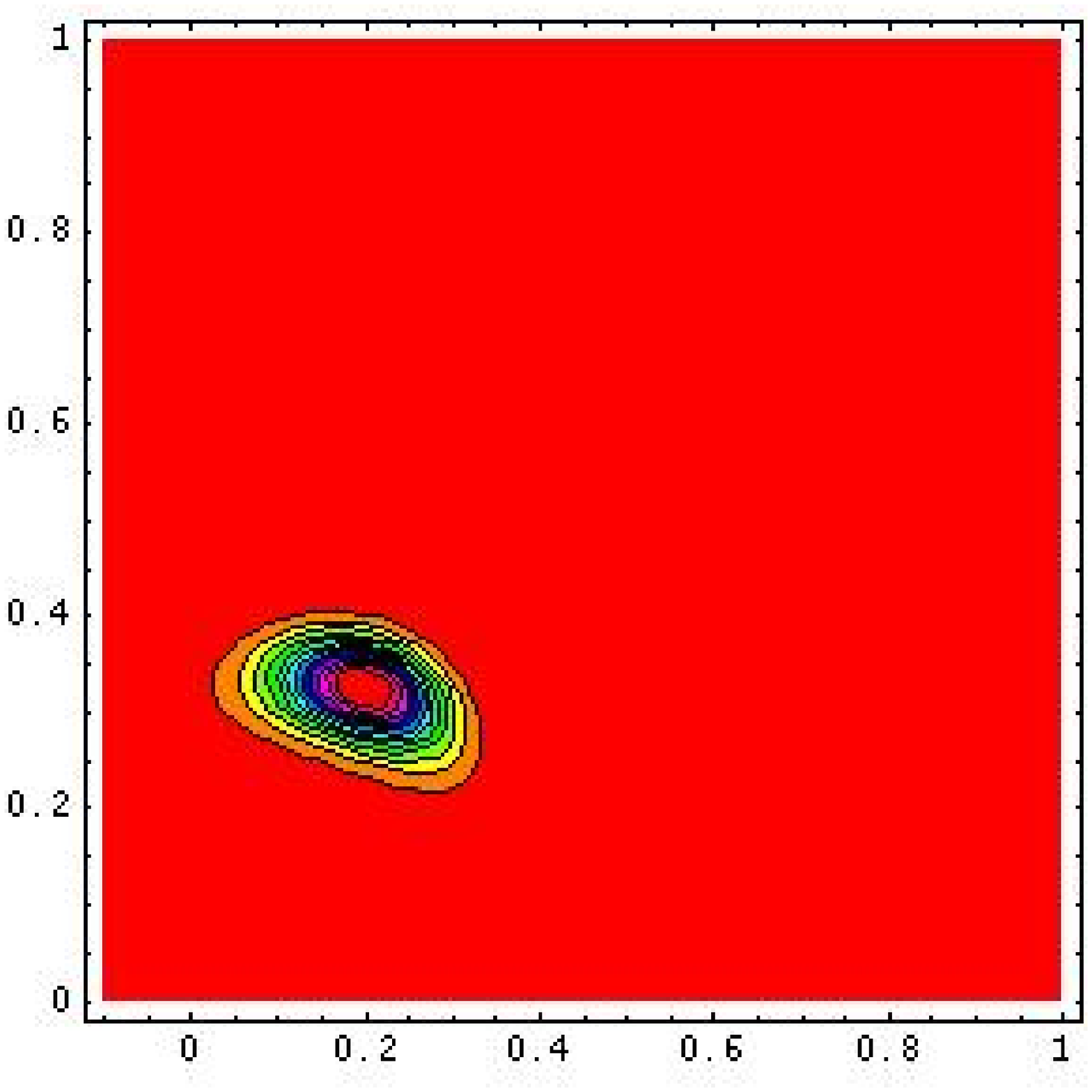,clip=,width=12.0cm} \\
\hline
\end{tabular}
\end{center}
\caption{\small \sl Probability density function and 
contour plot  
$f(\rhobar, \etabar\,|\,C1,C2,C3)$
obtained by the constraint given by 
$\vubovcb$, $\etabar$ and $\dmd$ (see remarks in text and in caption of
Fig. \ref{fig:fC1contour} about the interpretation of the contour plot).}
\label{fig:fC1C2C3}
\end{figure}

With this third reweighting, the resulting p.d.f. looses 
finally sign ambiguities, 
and becomes rather narrow, with respect to the initial space of possibilities.
The 3-D plot is shown in Fig. \ref{fig:fC1C2C3}.
At this point we can evaluate expected value and standard uncertainty 
of the quantities of interest:
\begin{equation}
f(\rhobar, \etabar\,|\,C1,C2,C3)\hspace{0.5cm} \Rightarrow \hspace{1.0cm} 
\begin{array}{l}
\left\{\begin{array}{l}  \mbox{E}[\rhobar]=0.19  \\
                         \sigma(\rhobar)=0.07 
\end{array}\right. \\
\mbox{} \\
\left\{\begin{array}{l} \mbox{E}[\etabar]=0.32 \\
                         \sigma(\etabar)=0.04 
\end{array}\right.                      
\end{array}
\label{eq:res1}
\end{equation}

\newpage

\section{Including the experimental information about $\dms$}
\begin{figure}
\begin{center}
\begin{tabular}{|c|}
\hline
\epsfig{file=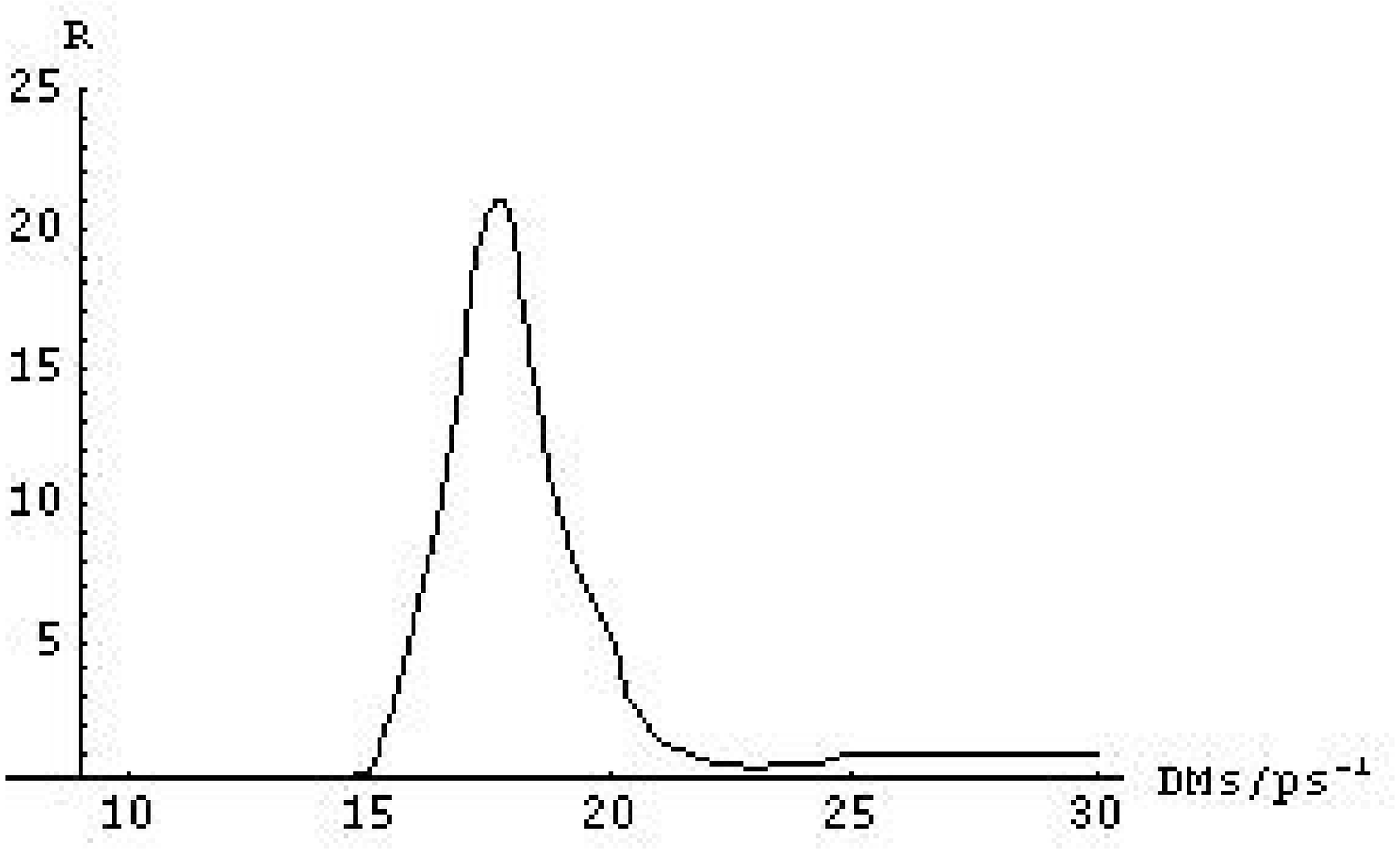,clip=,width=10.5cm,height=5.5cm} \\
\hline
\epsfig{file=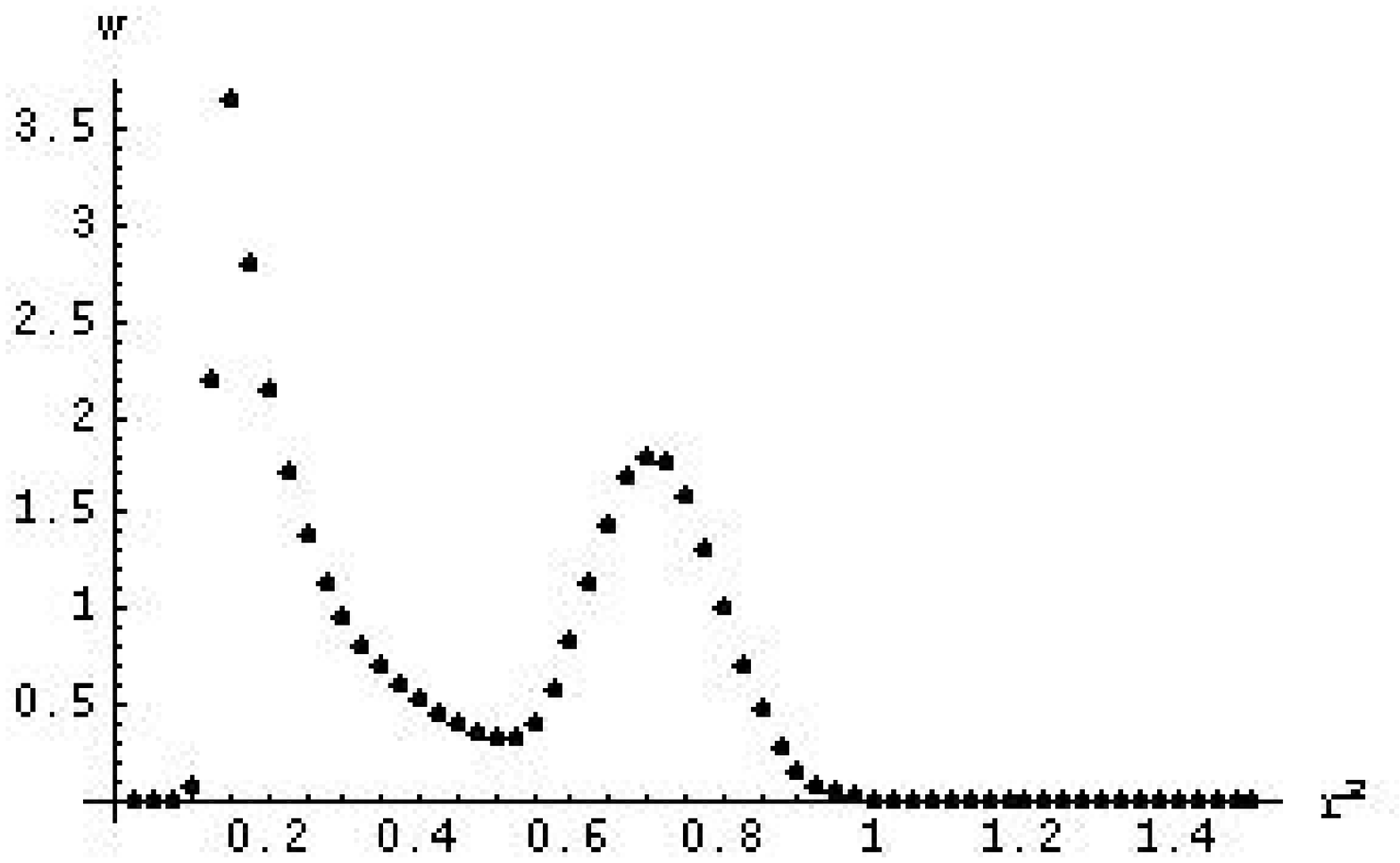,clip=,width=10.5cm,height=5.5cm} \\
\hline
\epsfig{file=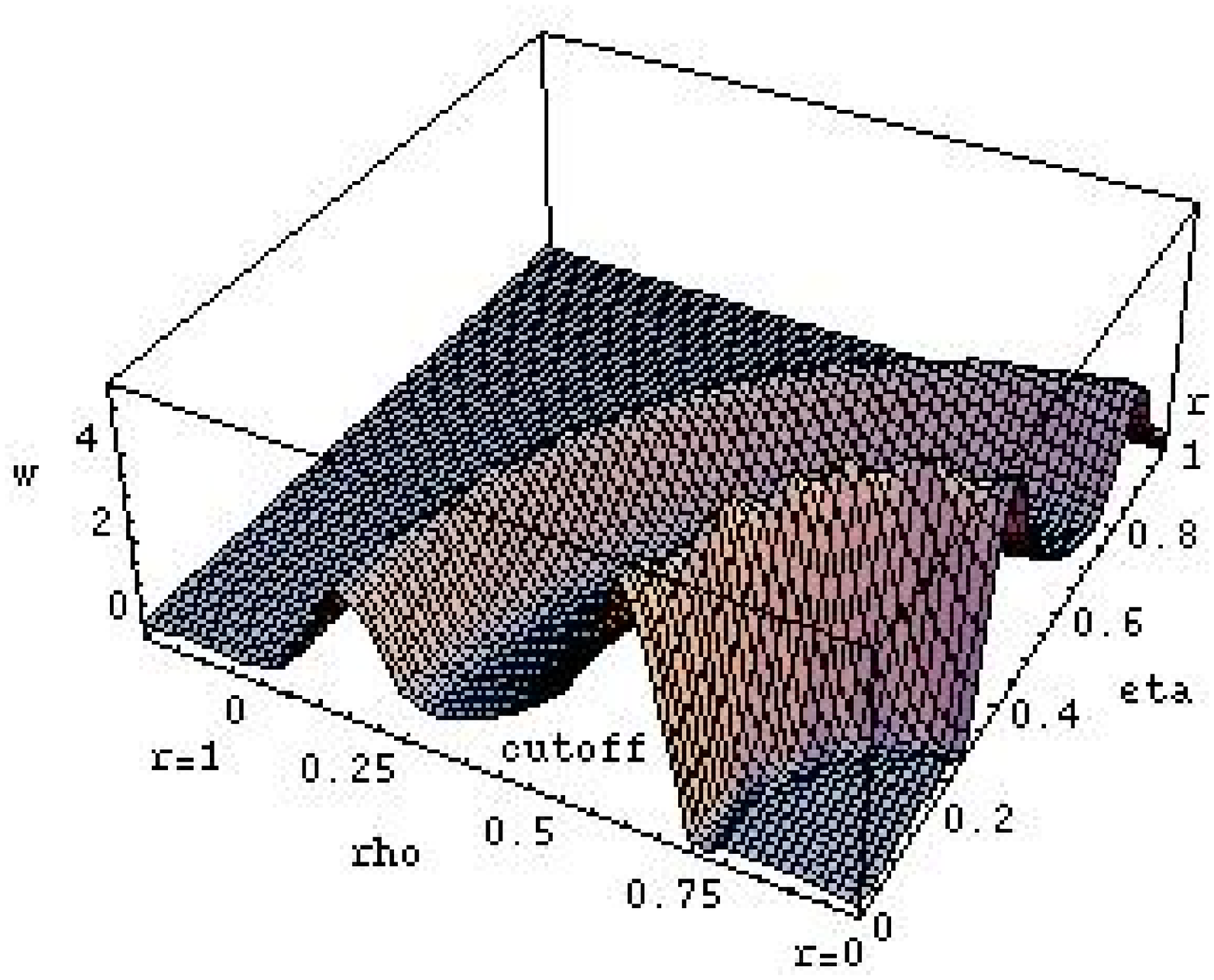,clip=,width=12.0cm} \\
\hline
\end{tabular}
\end{center}
\caption{\small \sl Top plot: likelihood of $\Delta m_s$
rescaled to the region of insensitivity. Central plot: reweighting 
factor of $r^2=e/\Delta m_s$. The peak just below
0.2 is an artifact of the  cutoff in $\dms$ (see text). 
Bottom plot: same   reweighting  
factor in the $\rhobar$--$\etabar$ plane (note also here the 
low $r$ cutoff).}
\label{fig:Rdms}
\vspace{-13.8cm}\mbox{} \hspace{3.5cm} {\bf \large cutoff}
\vspace{+13.8cm} 
\end{figure}

The treatment of the experimental information about 
$\dms$ cannot be done by the usual uncertainty propagation, just
because the simplifying hypotheses (linearization and 
Gaussian model for $e/\dms$) do not hold. 
In fact, the likelihood, which has the role of reweighting 
the probability, is {\it open}, in the sense described 
in Section 7 of Ref.~\cite{clw}, i.e. it does not go to zero 
at both ends of the kinematical region ($\dms=0$ and $\dms=\infty$
in this case), as shown in the top
plot of Fig. \ref{fig:Rdms}.
The reason is simple: in this kind of measurement
$\dms$ is not yet incompatible with $\infty$ (no oscillation). 
Though the open likelihood does not allow to renormalize the p.d.f.
(unless strong priors forbidding high values are used), 
the likelihood can still be used to reweigh the points in the
$\rhobar$--$\etabar$ plane (see Refs. \cite{conPeppe} and 
\cite{conPia} for other examples and discussions about the $R$ function).

It is instructive to see how the reweighting of $\dms$ is 
turned into the reweighting of the square radius of circle centered
in $\rhobar=1$ and $\etabar=0$ given by the constraint $C_4$, 
i.e. $r^2 = e/\dms$. This is illustrated in the central plot of 
Fig. \ref{fig:Rdms}. We see that the reflection of $R\rightarrow 0$
for $\dms < 14$ ps$^{-1}$ essentially kills  values above 
$r = 1$, resulting in a strong sensitivity bound (in the sense
of Ref.~\cite{clw}) on the angle $\gamma$ of the CKM matrix, 
forced to be
below 90$^\circ$. The bottom plot of Fig.  \ref{fig:Rdms} shows the
reweighting function in the $\rhobar$--$\etabar$ plane. 

For small values of $r^2$ the reweighting function is
divergent, since the whole region of high $\dms$ 
is squeezed into a small region of $r^2$. This is no serious 
problem, since these points are already ruled out by the other 
constraints. The cutoff shown in the central and bottom plot 
of Fig.  \ref{fig:Rdms} is due to a cutoff at $\dms=100$\,ps$^{-1}$. 
Note, moreover, that the values of $\dms$  preferred by the data 
(around 15--20 ps$^{-1}$) overlap well with the 
$\rhobar$--$\etabar$ region indicated by the other constraints. 
Note also that even if one goes through the  {\it academic exercise} 
of removing by 
hand the peak around  15--20 ps$^{-1}$, chopping $R$ to 1, 
the effect on the values
of $r$ above 1 (and hence of  $\gamma$  above 90$^\circ$) does 
not change. 
 
\section{Global combination and results}
\begin{figure}
\begin{center}
\begin{tabular}{|c|}
\hline
\epsfig{file=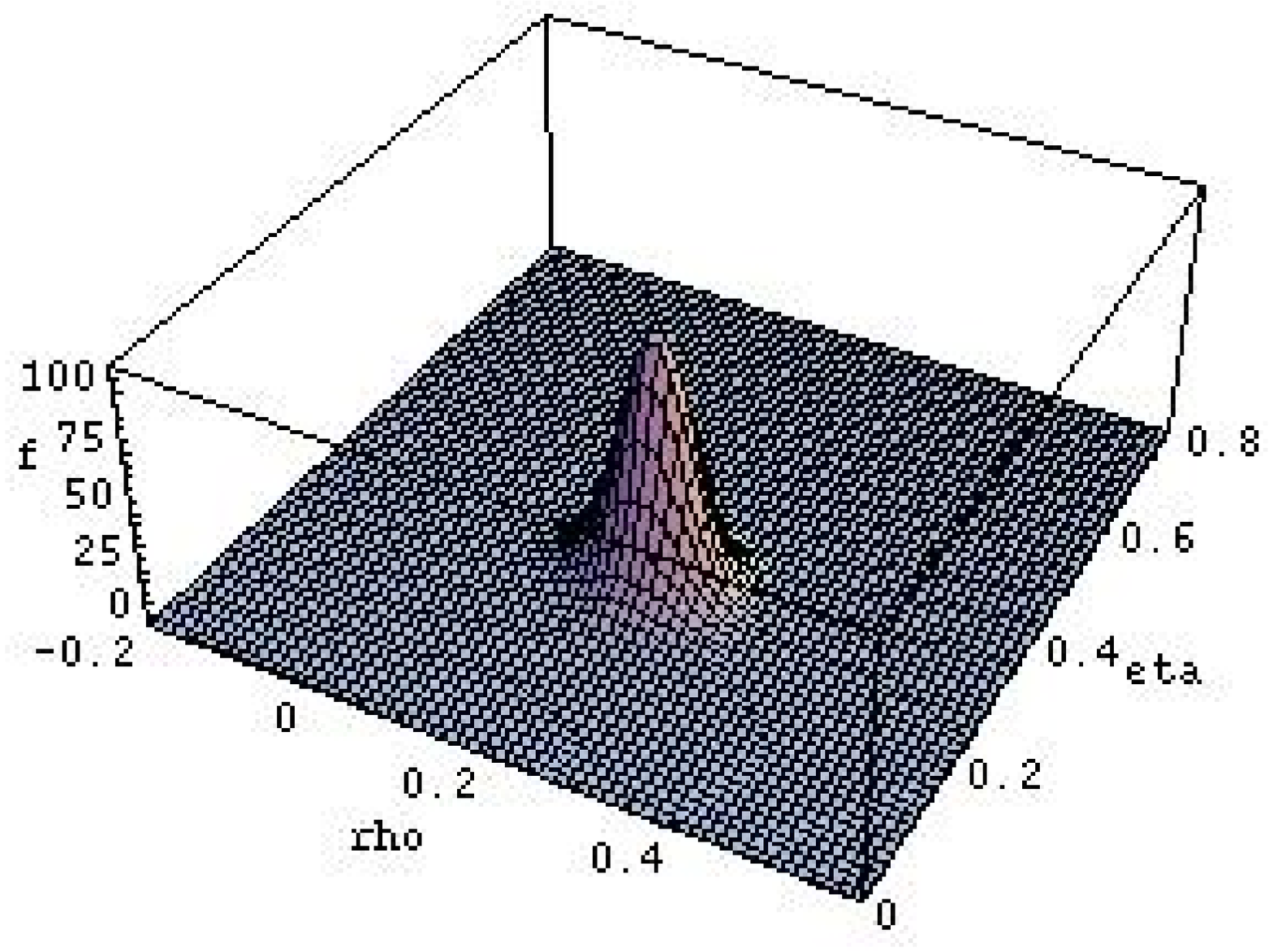,clip=,width=12.0cm} \\
\hline
\epsfig{file=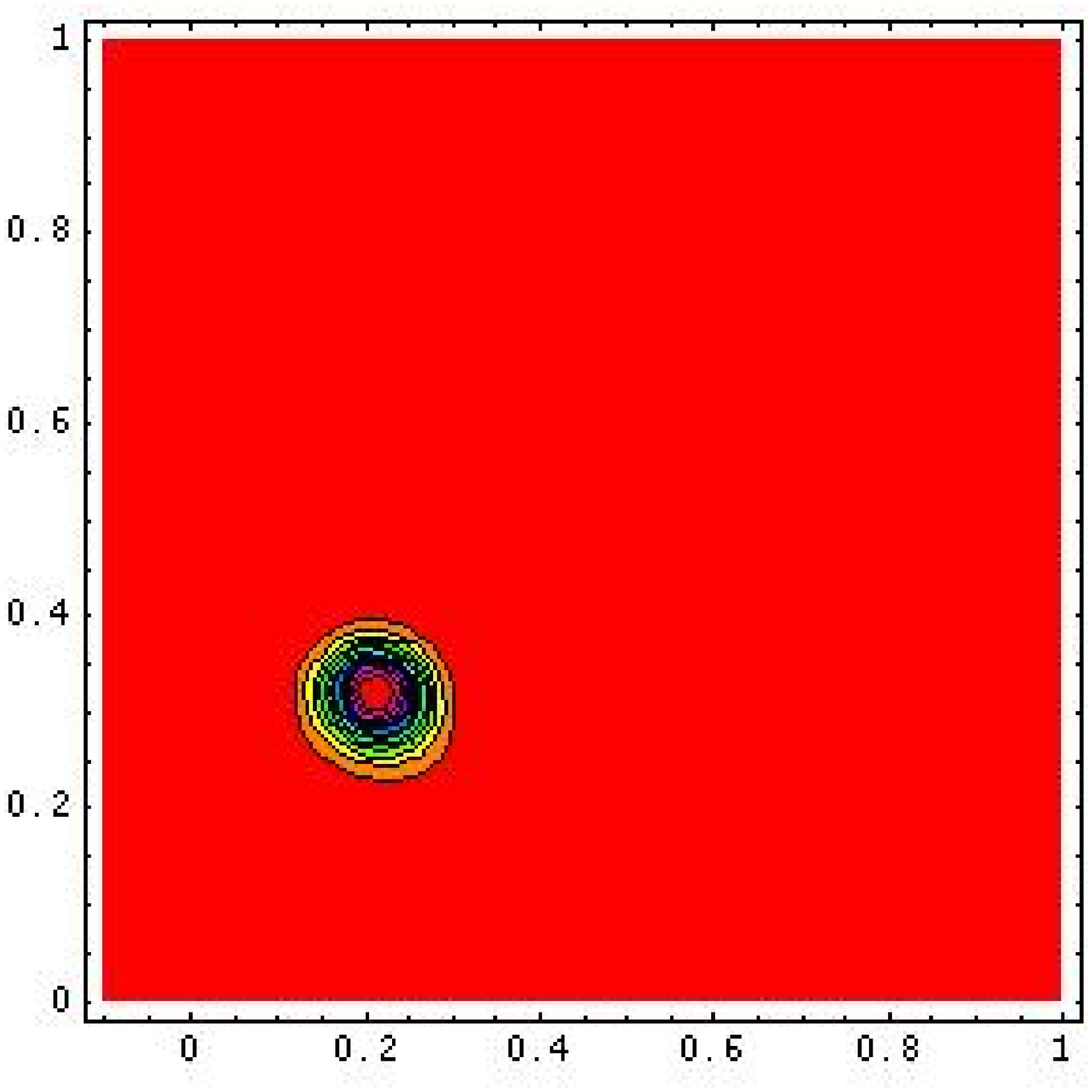,clip=,width=12.0cm} \\
\hline
\end{tabular}
\end{center}
\caption{\small \sl Probability density function and 
contour plot  
$f(\rhobar, \etabar\,|\,C1,C2,C3,C4)$
obtained by the constraint given by 
$\vubovcb$, $\etabar$, $\dmd$ and $\dms$ 
(see remarks in the text and in caption of
Fig. \ref{fig:fC1contour} about the interpretation of the contour plot).}
\label{fig:fAll}
\end{figure}

Reweighting  $f(\rhobar, \etabar\,|\,C1,C2,C3)$ with 
$R$ function of $\dms$ we get the final result shown in Fig.
\ref{fig:fAll}. 

Expected values and standard deviations are obtained by numerical
integration. The result is 
\begin{equation}
f(\rhobar, \etabar\,|\,C1,C2,C3,C4)\hspace{0.5cm} \Rightarrow \hspace{1.0cm} 
\begin{array}{l}
\left\{\begin{array}{l}  \mbox{E}[\rhobar]=0.213  \\
                         \sigma(\rhobar)=0.042
\end{array} \right. \\
\mbox{} \hspace{3.5cm} \rho(\rhobar,\etabar) = -0.12 \\
\left\{\begin{array}{l} \mbox{E}[\etabar]=0.314 \\
                         \sigma(\etabar)=0.039
\end{array}\right.                      
\end{array}
\label{eq:res2}
\end{equation}
practically identical to 
$\rhobar=0.224\pm 0.038$  and  
$\etabar=0.317\pm 0.040$ given in Ref.~\cite{GOR} using the full 
Monte Carlo integration (obviously, those who pay attention the tenths 
of standard deviations should use the more accurate result of 
Ref.~\cite{GOR}).

\begin{figure}
\begin{center}
\begin{tabular}{c}
\epsfig{file=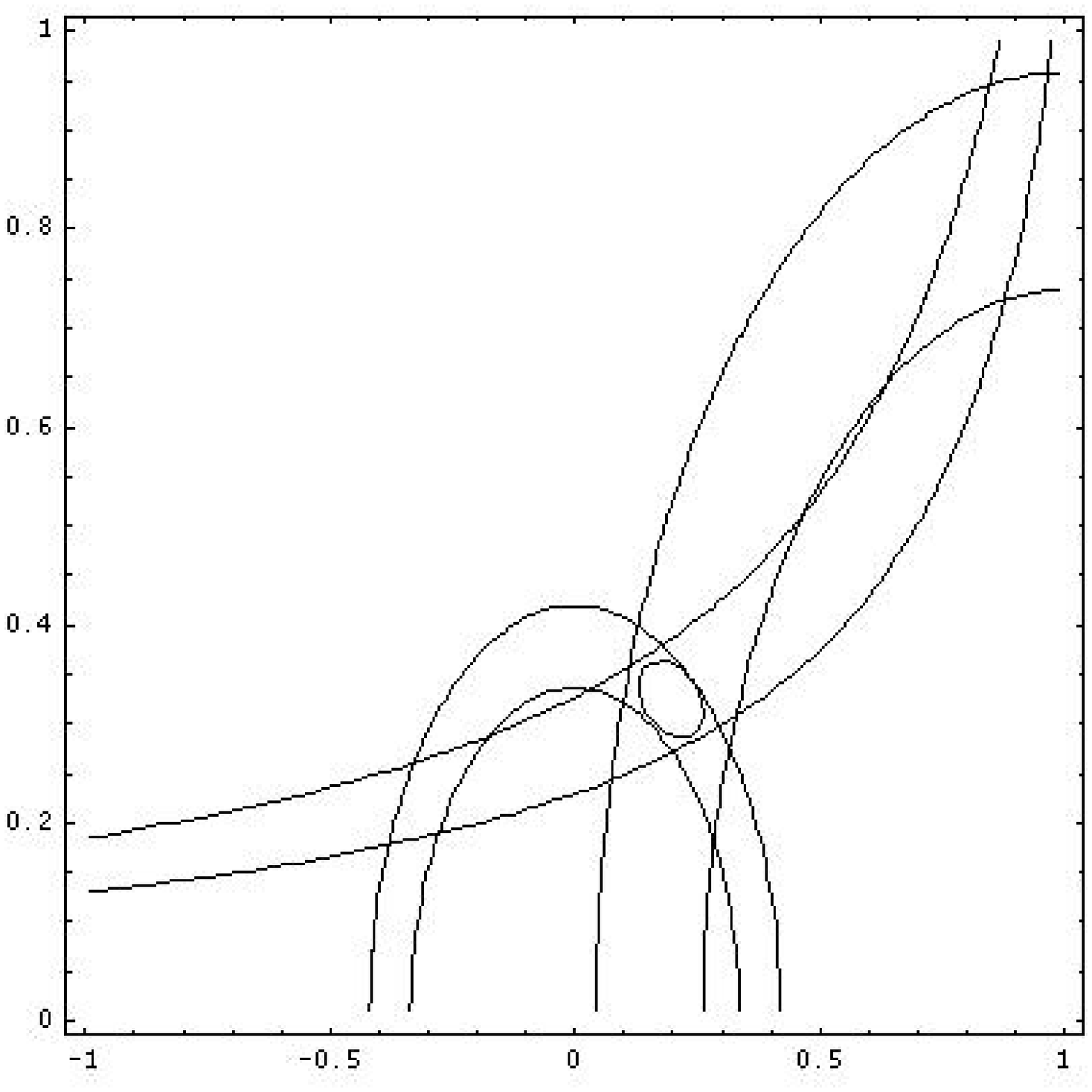,clip=,width=10.0cm} \\
\epsfig{file=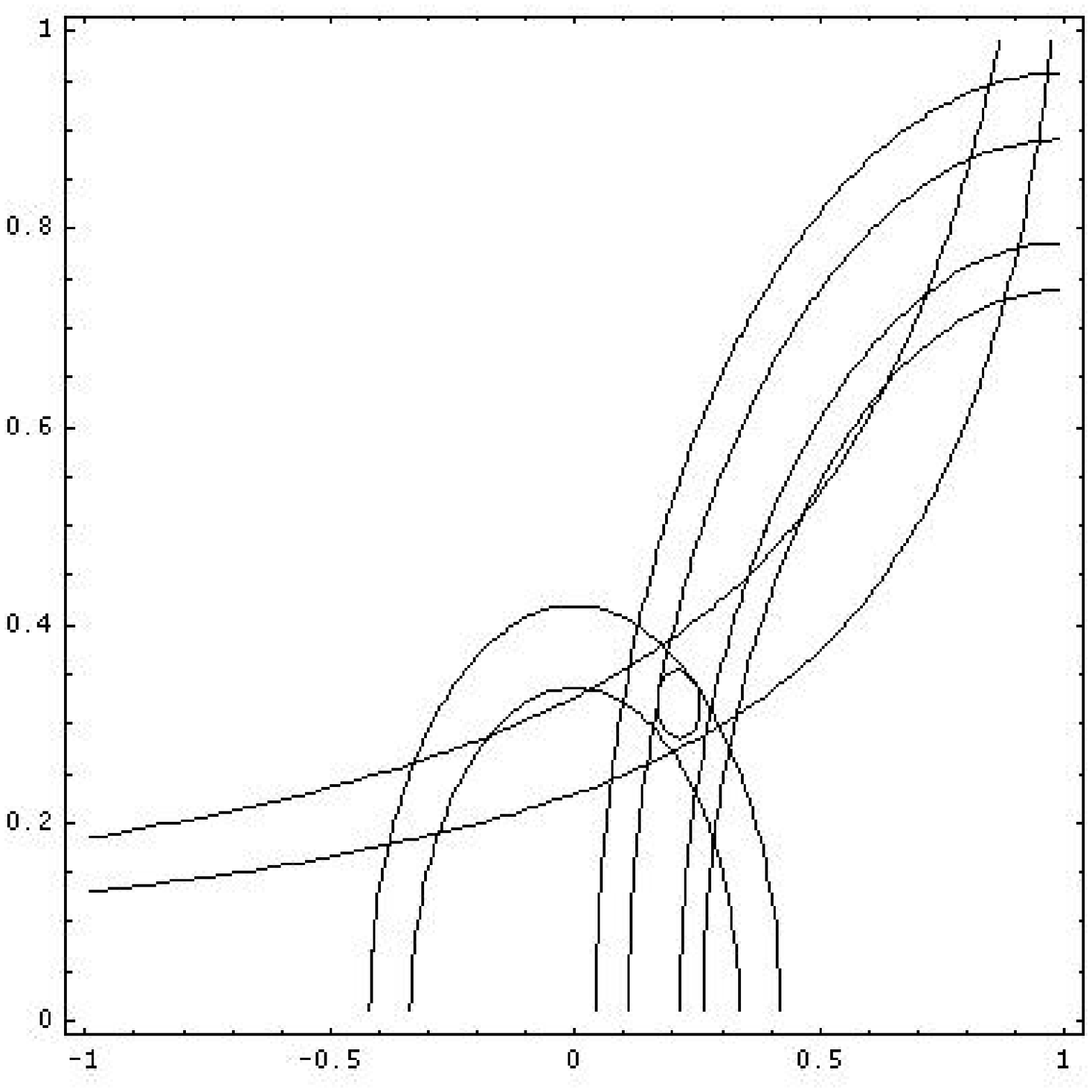,clip=,width=10.0cm} \\
\end{tabular}
\end{center}
\caption{\small \sl Top plot: 
$\Delta (\ln L)=1/2$ contours from the constraints
$C_1$, $C_2$, $C_3$ and their combination. 
The bottom plot shows the effect of the 
experimental information on $\dms$, limited to the region 
of interest.  The probability that $\rhobar$ and 
$\etabar$ are \underline{both} in the ``ellipses'' is
about 37\%.}
\label{fig:contours}
%\vspace{-11.5cm}\hspace{2.4cm} $\etabar$ \\
\vspace{-21.5cm}\hspace{2.4cm} $\etabar$ 
\\ \\ \\ \\ \\ \\ \\ \\ \\ \\ \\ 
\mbox{}\hspace{4.2cm}$\epsilonk$
\\ \\ \\ \\ 
\mbox{}\hspace{5.4cm}$\vubovcb$
\hspace{2.18cm}$\dmd$
\\ \\ \\ \\
\mbox{}\hspace{2.4cm} $\etabar$\\
%\vspace{+10cm}
\\ \\ \\ \\ 
\mbox{}\hspace{10.cm}$\dms$
\\ \\ \\ \\ \\ \\ \\ \\ \\ \\  \\ \\ \\
\mbox{}\hspace{11.8cm} $\rhobar$\\ \\ \\ \\
\end{figure}

It is interesting to show partial and global results 
as contour lines at $e^{-1/2}=0.61$ of the maximum of the 
reweighting functions, equivalent to the $\Delta (-\ln{L})=1/2$ or 
 $\Delta\chi^2=1/2$ rules\footnote{Note that the translation 
of these rules into ``confidence'' is not straightforward, 
unless the log--likelihood (or the $\chi^2$) has a nice 
parabolic shape and all values where the likelihood
finally concentrates are initially considered about equally likely.}
(I refer to Ref.~\cite{GOR} for the relation between 
``standard'' methods based on $\chi^2$ minimization 
and the more detailed inferential scheme illustrated there). 
The top plot of Fig.~\ref{fig:contours} shows the contour
``roads'' 
given by the first three constraints, together with 
the (almost) ellipse of their combination. The probability 
that the % true 
values of $\rhobar$ and $\etabar$ are 
\underline{both} in the ellipse is about 
37\%.\footnote{This value should be compared with 39.3\% 
obtainable in the case of a perfect 2-D Gaussian, thus indicating 
that the final distribution comes out \underline{naturally} 
approximatively of that kind.} 
Instead, the projections of the ellipse on each axis gives 
an interval of about $68\%$ probability  in \underline{each}
variable.

The bottom plot of Fig.~\ref{fig:contours} shows the effect
of the constraint $C_4$. First we notice the perfect agreement
between the $\dmd$ and $\dms$ roads, indicating that the 
values of $\dms$ suggested by the data are absolutely 
consistent with the other constraints within the Standard Model. 
Furthermore, the effect of the $\dms$ on the ``ellipse'' of the 
final inference is to reshape the left side, increasing the value
of $\rhobar$ and decreasing its uncertainty, with almost no 
effect on $\etabar$, as also shown by the results  (\ref{eq:res1})
 and (\ref{eq:res2}). 

\section{Conclusion}
The result of Ref.~\cite{GOR} has been reproduced, using
some simplifying assumptions that allow most calculations
to be performed analytically. The consistency of the pieces information
on $\rhobar$ and $\etabar$ coming from the different constraints, 
as well as that relating the constraints to the uncertain parameters 
of the theory (see Ref.~\cite{GOR} for details), is almost ``too good'', 
suggesting that the final uncertainties are most likely 
somewhat \underline{over}estimated. 

We all are looking forward to the results of the $B$ factories
and the next generation of experiments  on 
the CKM matrix parameters,  hoping to find 
evidence of new phenomenology that would give new vitality 
to particle physics. 
\vspace{4.0mm}

It is a pleasure to thank my friends Achille et al. (Ref. \cite{GOR})
which have dragged me into this business.  
%\section*{Acknowledgements}
%It is a pleasure to thank Achille, Enrico, Fabrizio, Guido, Marco, 
%Patrik and Vittorio who have dragged me into this business. 
%Indeed it has been a nice opportunity to rethink data analysis issues,
%and to learn for the first time 
%some of the physics issues related to Ref.~\cite}. However, 
%entering for a while in the ``triangulation polemics''
% lead me also to sad considerations about sociology 
%of science. 

%\vspace{1.0cm}

\end{document}